# Resonant mode conversion in the waveguides with an unbroken and broken $\mathcal{PT}$-symmetry

Victor A. Vysloukh[1,*] and Yaroslav V. Kartashov[2,3]

[1]*Departamento de Fisica y Matematicas, Universidad de las Americas-Puebla, Santa Catarina Martir, 72820 Puebla, Mexico*
[2]*ICFO-Institut de Ciencies Fotoniques, and Universitat Politecnica de Catalunya, Mediterranean Technology Park, 08860 Castelldefels (Barcelona), Spain*
[3]*Institute of Spectroscopy, Russian Academy of Sciences, Troitsk, Moscow Region 142190, Russia*

We study resonant mode conversion in the $\mathcal{PT}$-symmetric multimode waveguides, where symmetry breaking manifests itself in sequential destabilization (appearance of the complex eigenvalues) of the pairs of adjacent guided modes. We show that the efficient mode conversion is possible even in the presence of the resonant longitudinal modulation of the complex refractive index. The distinguishing feature of the resonant mode conversion in the $\mathcal{PT}$-symmetric structure is a drastic growth of the width of the resonance curve when the gain/losses coefficient approaches a critical value, at which symmetry breaking occurs. We found that in the system with broken symmetry the resonant coupling between exponentially growing mode with stable higher-order one effectively stabilizes dynamically coupled pair of modes and remarkably diminishes the average rate of the total power growth.

In the end of nineties it was established that non-Hermitian Hamiltonians satisfying a specific parity-time symmetry condition can possess purely real eigenvalue spectrum - an indispensable attribute of any stable physical quantum system [1]. Optical systems, such as couplers, built as the balanced pairs of amplifying and absorbing waveguides, serve as analogs of quantum systems with $\mathcal{PT}$-symmetric potentials and are capable of supporting guided modes whose power remains constant upon propagation [2]. The total power conservation can be established only due to the presence of the transverse energy flow from the amplifying waveguide to the absorbing one, provided that the amplitude of gain/losses does not exceed the critical value, at which symmetry breaking occurs. Experimental evidences of this important property are available [3,4]. Nowadays linear as well as nonlinear static $\mathcal{PT}$-symmetric systems are extensively studied in various configurations. Switching operations, unidirectional dynamics, and nonreciprocal soliton scattering were reported [5-10]. These salient features are linked with such intriguing application as a unidirectional invisibility [11,12]. Purely nonlinear variants of $\mathcal{PT}$-symmetric structures were reported too [13,14].

An important new branch of this activity is connected with $\mathcal{PT}$-symmetric structures with the longitudinally-modulated parameters. This setting attracts attention due to rich physics and a variety of applications. The possibility of dynamic localization of a wave packet due to harmonic bending was illustrated in the photonic lattice with an unbroken $\mathcal{PT}$-symmetry, while highly nonreciprocal Bragg-scattering was shown to occur at the breaking point [15]. $\mathcal{PT}$-symmetric coupler with the longitudinal modulation of the coupling constant proved to be analogous to the parametric oscillator exhibiting an unusual behavior [16]. High-frequency longitudinal modulation of the complex potential can drive the transition between broken-$\mathcal{PT}$ and unbroken-$\mathcal{PT}$ phases [17]. The idea of Kapitza stabilization was extended to the case of the imaginary single-well oscillating potential [18]. The possibility to manipulate the pseudo-$\mathcal{PT}$-symmetry by applying a periodic modulation of the complex refractive index was suggested in [19]. Finally, the stochastic modulation of the parameters of $\mathcal{PT}$-symmetric coupler with balanced, on average, gain and losses was considered in [20] and it was shown that the intensity of the field grows independently of the type of fluctuations.

Notice that all this research activity was focused on couplers and lattices consisting of the simplest single-mode waveguides. However, structures involving multimode waveguides provide additional degrees of freedom. For instance, in the longitudinally modulated Hermitian (conservative) multimode waveguides, one can observe [21,22] stimulated transitions between confined light modes akin to Rabi-like flopping in the multilevel quantum systems.

In this Letter we discuss specific features of the resonant mode conversion in multimode waveguides with an unbroken and broken $\mathcal{PT}$-symmetry. We analyze the trajectories of the complex eigenvalues upon the increase of the amplitude of gain/losses and then focus our attention on the resonant properties and efficiency of conversion of modes stimulated by the harmonic longitudinal modulation of the refractive index and amplification in the regime of unbroken $\mathcal{PT}$-symmetry. In the case of broken $\mathcal{PT}$-symmetry we discuss the possibilities for suppression of exponential growth of the power by the resonant dynamical mode binding.

We consider the propagation of a light beam along the $\xi$-axis of a multimode waveguide governed by the nonlinear Schrödinger equation for dimensionless field amplitude $q$:

$$i\frac{\partial q}{\partial \xi} = -\frac{1}{2}\frac{\partial^2 q}{\partial \eta^2} - R(\eta,\xi)q. \qquad (1)$$

Here, the longitudinal $\xi$ and transverse $\eta$ coordinates are scaled to the diffraction length and the input beam width [21]; the complex function $R(\eta,\xi) = U(\eta,\xi) + iV(\eta,\xi)$ describes the "potential" affecting the propagation of the laser radiation. Its real part describes the refractive index profile that is symmetric in the transverse direction and is harmonically modulated with the frequency $\Omega$ and the amplitude $\mu \ll 1$ in the longitudinal direction:

$U(\eta,\xi) = p_r[1+\mu\sin(\Omega\xi)]/\cosh(\eta)$. The parameter $p_r$ stands for the waveguide depth. The imaginary part, responsible for the amplification and absorption, is antisymmetric: $V(\eta,\xi) = p_i[1+\mu\sin(\Omega\xi+\phi)]\tanh(\eta)/\cosh(\eta)$, where $p_i$ parameter stands for the amplification coefficient, while $\phi$ describes the possible phase shift between the harmonic modulation of the refractive index and gain/losses.

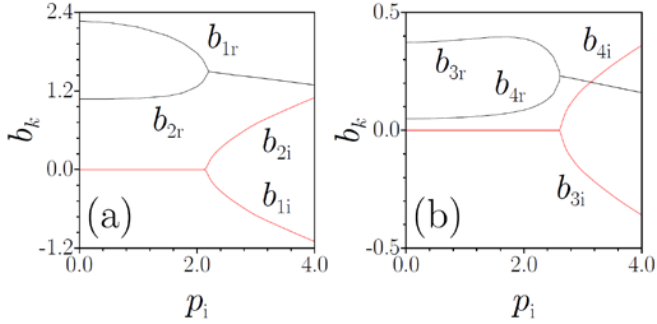

Fig. 1. Real and imaginary parts of the propagation constants of the first and second (a), as well as of the third and fourth (b) eigenmodes versus the amplitude of gain and losses $p_i$. The collision of eigenvalues $b_{1r}$ and $b_{2r}$ occurs at $p_i = 2.16$, while $b_{3r}$ and $b_{4r}$ collide at $p_i = 2.60$. In all cases $p_r = 3$.

For illustration of the dynamics of stimulated mode conversion the eigenmodes $w_m(\eta)$ of the complex waveguide and corresponding (generally also complex) propagation constants $b_m = b_{mr} + ib_{mi}$ were found numerically upon solution of the following linear eigenvalue problem $bw = (1/2)d^2w/d\eta^2 + R|_{\xi=0}w$. Further we set $p_r = 3$ that corresponds to the situation, when in the absence of gain/losses ($p_i = 0$) the waveguide supports only four guided modes (see Figs. 1 and 2 where their properties are summarized). If $p_i$ grows, the real parts $b_{mr}$ of propagation constants of the neighboring modes of different parity approach each other, i.e. $b_{1r} \leftrightarrow b_{2r}$ and $b_{3r} \leftrightarrow b_{4r}$. The collision of the eigenvalues first occurs at $p_i \approx 2.16$ for the first and second modes. Further increase of $p_i$ is accompanied by the emergence of two complex asymmetric modes with propagation constants having equal real parts, but opposite imaginary parts [Fig. 1(a)]. The collision of the eigenvalues for the second pair of modes ($w_3, w_4$) occurs for higher gain amplitude $p_i \approx 2.60$ [Fig. 1(b)]. Corresponding transformation of the intensity distributions of different modes and their spatial spectra are illustrated in Fig. 2. The most important feature is growing with $p_i$ similarity of the intensity profiles of different modes [compare black and red curves in panels (a),(b) and notice the tendency for contrast reduction in the shape of second and third modes with increase of $p_i$]. These shape transformations are accompanied by the notable shift of the integral centers of spatial spectra $s(\omega) = \int w(\eta)\exp(i\omega\eta)d\eta$ to the low-frequency region, that indicates on the increase of the energy flow from the amplifying to the absorbing domain.

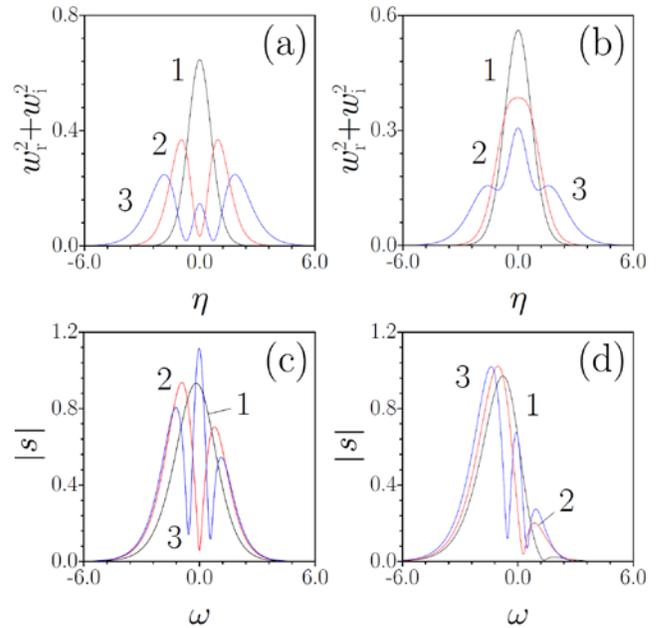

Fig. 2. Intensity distributions for the first three eigenmodes of $\mathcal{PT}$-symmetric wave-guide at $p_i = 0.5$ (a) and $p_i = 2.0$ (b). Panels (c),(d) show spectral amplitudes corresponding to (a),(b), respectively. In all the cases $p_r = 3$.

In the case of shallow longitudinal modulation of the complex refractive index ($\mu \ll 1$) the total field distribution can be approximately described by the superposition of the resonantly-coupled modes of equal parity $q(\eta,\xi) = c_1(\xi)w_1(\eta)e^{ib_1\xi} + c_3(\xi)w_3(\eta)e^{ib_3\xi}$. At the resonance condition $\Omega = |b_1 - b_3|$ the evolution of modal weights $c_k(\xi)$ is governed by the following equations:

$$\frac{dc_1}{d\xi} = i\mu\mathcal{I}_{11}\sin(\Omega\xi)c_1 - \frac{\mu}{2}(\mathcal{C}_{13} - \mathcal{S}_{31})c_3,$$
$$\frac{dc_3}{d\xi} = i\mu\mathcal{I}_{33}\sin(\Omega\xi)c_3 + \frac{\mu}{2}(\mathcal{C}_{13} + \mathcal{S}_{31})c_1, \quad (2)$$

with the exchange integrals given by

$$\mathcal{I}_{kk} = p_r\int_{-\infty}^{\infty}|w_k|U|w_k|d\eta,$$
$$\mathcal{C}_{km} = p_r\int_{-\infty}^{\infty}|w_k|U|w_m|\cos(\phi_m - \phi_k)d\eta, \quad (3)$$
$$\mathcal{S}_{km} = p_i\int_{-\infty}^{\infty}|w_k|V|w_m|\sin(\phi_m - \phi_k)d\eta.$$

The main difference of these coupled-mode equations with their "conservative" counterparts [21] stems from the nontrivial phase structure $w_k = |w_k|\exp[i\phi_k(\eta)]$ of the modes and the presence of the terms $\sim \mathcal{S}_{km}$ affecting the coupling strength. It can be shown using Eq. (2) that the actual frequency of energy exchange between coupled modes under resonant modulation $\Omega = |b_1 - b_3|$ is given by $\Omega_c = \mu(\mathcal{C}_{13}^2 - \mathcal{S}_{31}^2)^{1/2}$. It is proportional to the modulation amplitude $\mu$ and depends on the exchange integrals that account for the phase modulation of coupled modes as well as for the shapes of both refractive index and gain. Notice that upon derivation of Eqs. (2) the condition of the orthogonality of guided modes should be replaced by the biorthogonality condition, since $p_i \neq 0$ [23].

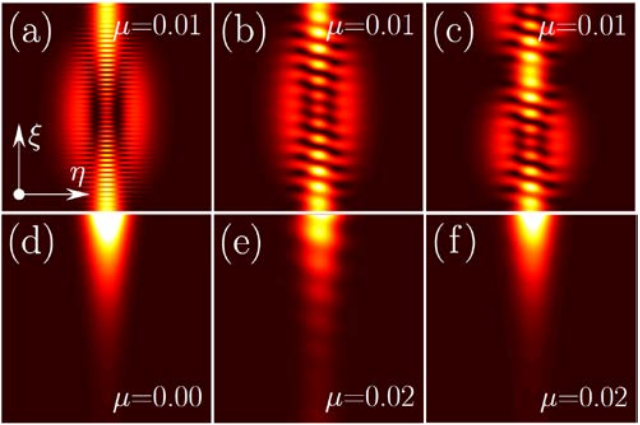

Fig. 3. Beam evolution in the longitudinally modulated $\mathcal{PT}$-symmetric waveguide at (a) $p_i=0$, $\mu=0.01$, (b) $p_i=1$, $\mu=0.01$, $\phi=0$, (c) $p_i=1$, $\mu=0.01$, $\phi=\pi$, (d) $p_i=2.18$, $\mu=0$, (e) $p_i=2.18$, $\mu=0.02$, $\phi=\pi$, and (f) $p_i=2.18$, $\mu=0.02$, $\phi=0$. Panels (a)-(c) correspond to the unbroken $\mathcal{PT}$ symmetry and propagation distance $\xi=1780$, while panels (d)-(f) correspond to the broken $\mathcal{PT}$ symmetry and propagation distance $\xi=40$. In all the cases only the first mode is launched into the waveguide at $\xi=0$ and the resonance condition is accomplished.

Direct numerical solution of Eq. (1) in the case when only fundamental mode $w_1$ is provided at the input clearly illustrates substantial differences in the dynamics of mode conversion (as an example we illustrate resonant conversion process $w_1 \to w_3$) in the conservative [$p_i=0$, Fig.3(a)] and $\mathcal{PT}$-symmetric [$p_i=1$, Figs. 3(b) and 3(c)] systems. While mode conversion in the $\mathcal{PT}$-symmetric system is still possible, the intensity distributions in the $(\eta, \xi)$ plane become somewhat "slant" due to the presence of the considerable transverse energy flows in the $\mathcal{PT}$-symmetric system. The period of small-scale intensity modulation in the longitudinal direction notably increases in the latter case. Importantly, the application of the out-of-phase modulation of the refractive index and gain/losses ($\phi=\pi$) drastically reduces the conversion length, i.e. the distance at which the energy weight $\nu_3=|c_3|^2$ of the third mode acquires its maximal value [compare panels 3(b) and 3(c)]. The impact of the phase shift $\phi$ on the conversion dynamics is even more visible in Figs. 4(a) and 4(b), where the evolution of the mode weights $\nu_k(\xi)=|c_k(\xi)|^2$ is depicted for the in-phase and out-of-phase refractive index and gain modulation. In fact one can distinguish three different scales in Figs. 4(a),(b) with smallest of them given by $2\pi/\Omega$. The frequency of somewhat slower oscillations does not depend notably on $\phi$ and is connected with the frequency of oscillation of slightly perturbed $w_1$ eigenmode of the waveguide. Finally, the frequency of slowest and deepest weight oscillations depends on the global mode coupling strength, e.g. on $\mu$, as it follows from the simple analytical approach [Eq. (2)].

As in the conservative case [21], the key factor determining the efficiency of mode conversion in the $\mathcal{PT}$-symmetric system is the detuning $\delta\Omega=\Omega-\Omega_{13}$ between the actual frequency of the refractive index modulation and the intrinsic mode beating frequency $\Omega_{13}=|b_1-b_3|$. Figure 5(a) shows a typical resonance curve, i.e. the dependence of the maximal weight of the third mode $\nu_3^{\max}$ on the frequency detuning $\delta\Omega$, for the case of $p_i=1$, when the system still remains $\mathcal{PT}$-symmetric. The width of this resonance curve is determined, among other factors, by the $\mu$ parameter and it rapidly decreases as $\mu \to 0$.

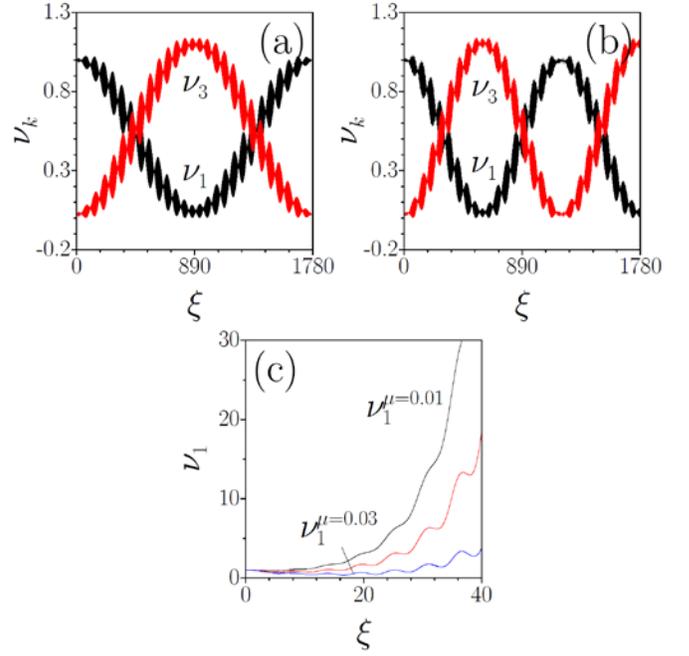

Fig. 4. The evolution of mode weights in the longitudinally modulated waveguide at (a) $p_i=1$, $\mu=0.01$, $\phi=0$, (b) $p_i=1$, $\mu=0.01$, $\phi=\pi$, and (c) $p_i=2.18$, $\phi=\pi$, and $\mu=0.01$, $0.02$, and $0.03$. In all cases only the first mode is launched into the waveguide at $\xi=0$.

The central result of this Letter is a drastic growth of the resonance width that takes place when the amplification parameter approaches the critical value $p_i \approx 2.16$ at which the first symmetry breaking occurs [Fig. 5(b)]. The reason behind such rapid resonance broadening is growing similarity of mode shapes with increase of $p_i$ (see Fig. 2 where it is obvious how multi-pole structure of higher-order modes is gradually lost), as well as rapid diminishing of the propagation constant difference between first and second modes. Notice that the second mode at low $p_i$ values is initially out of resonance and definitely not involved into conversion process due to its symmetry. However, since increasing $p_i$ results in substantial mode shape transformation, the corresponding overlap integrals become nonzero and considerable fraction of power can be transferred also into second mode, that now couples to both first and third modes. This expands the range of frequencies where power effectively transfers into third mode and leads to expansion of the resonant frequency band. The broadening of the resonant curve is accompanied by the progressively increasing amplitude of energy weight oscillations. Figure 5(c) illustrates the role of the phase shift $\phi$ between refractive index and gain modulation and confirms that the conversion length $\xi_{\text{tr}}$ acquires its minimal value at $\phi=\pi$, when modulation is out-of-phase. Notice that the conversion length monotonically decreases with the increase of the gain strength $p_i$ [Fig. 5(d)].

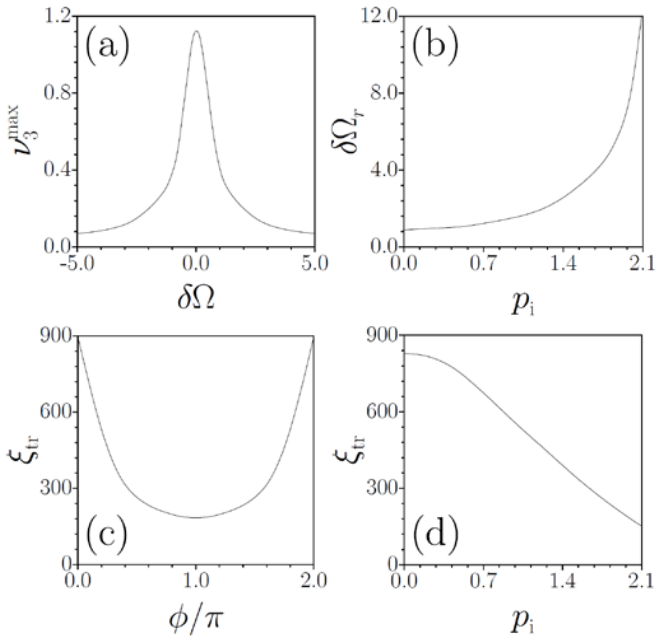

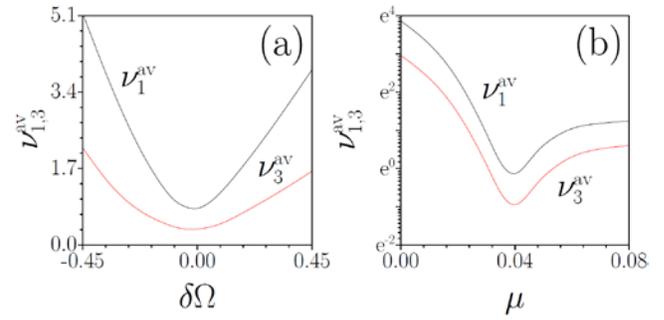

Fig. 5. (a) Maximal weight of the third mode versus frequency detuning $\delta\Omega$ at $p_i = 1$. (b) The width of the resonance curve versus $p_i$. In panels (a),(b) the out-of-phase modulation of the refractive index and gain-losses is used. (c) Coupling distance between first and third modes versus phase difference $\phi$ at $p_i = 2$. (d) Coupling distance versus $p_i$ for $\phi = \pi$. In all cases $p_r = 3$ and $\mu = 0.01$.

An intriguing scenario, that is in principle impossible in the case of a single-mode waveguide, is realized when the exponentially growing mode in the system with broken $\mathcal{PT}$-symmetry is resonantly coupled to the higher-order mode for which symmetry breaking occurs at larger gain value $p_i$ and that still has purely real propagation constant (i.e. the mode does not decay or grow upon evolution). Figure 3(d) shows fast exponential growth of the unstable first mode in the absence of coupling $(\mu = 0)$. This growth is drastically diminished, although not eliminated completely, if this mode resonantly couples to the third stable mode of the system at $\mu = 0.02$ and $\phi = \pi$ [Fig. 3(e)]. Thus, Fig. 4(c) illustrates evolution of growing weights of the first mode $\nu_1$ for different values of the modulation amplitude $\mu$. The stabilizing action of stronger mode coupling is clear. To quantify this process we introduced the distance-averaged energy weights $\nu_k^{av} = L^{-1}\int_0^L \nu_k(\xi)d\xi$. Expectably, the distance-averaged energy weights reach their minima under the condition of the exact resonance $\delta\Omega = 0$ [Fig. 6(a)], but the dependence of $\nu_k^{av}$ on the refractive index modulation depth $\mu$ is nontrivial and it shows a local minimum at certain intermediate value of the modulation depth $\mu$ [Fig. 6(b)]. Notice that too large modulation depth $\mu$ does not lead to further suppression of power growth, but causes pronounced radiative losses. It should be also stressed that only the out-of-phase modulation of the refractive index and gain allows to decrease power growth.

In summary, we showed that efficient mode conversion is possible in the $\mathcal{PT}$-symmetric systems, provided that the $\mathcal{PT}$-symmetry is not broken. We found that the resonance width drastically increases upon approaching the symmetry-breaking point. We also found that in the regime of broken $\mathcal{PT}$-symmetry the resonant coupling of the unstable exponentially growing mode with still stable higher-order one stabilizes this dynamically-bounded pair, by remarkably diminishing the average power growth rate.

Fig. 6. Distance-averaged mode weights as functions of (a) detuning $\delta\Omega$ at $\mu = 0.04$, and (b) modulation depth $\mu$ at $\delta\Omega = 0$. In both cases $p_i = 2.18$.